\documentclass[twocolumn,showpacs,superscriptaddress,prl]{revtex4}
\usepackage[latin9]{inputenc}
\usepackage{color}
\usepackage{verbatim}
\usepackage{amsmath}
\usepackage{amssymb}
\usepackage{graphicx}
\usepackage{esint}
\usepackage{hyperref}

\makeatletter
\@ifundefined{textcolor}{}
{%
 \definecolor{BLACK}{gray}{0}
 \definecolor{WHITE}{gray}{1}
 \definecolor{RED}{rgb}{1,0,0}
 \definecolor{GREEN}{rgb}{0,1,0}
 \definecolor{BLUE}{rgb}{0,0,1}
 \definecolor{CYAN}{cmyk}{1,0,0,0}
 \definecolor{MAGENTA}{cmyk}{0,1,0,0}
 \definecolor{YELLOW}{cmyk}{0,0,1,0}
}

\usepackage{mathrsfs}

\draft
\makeatother

\begin{document}

\title{Manipulating Topological Edge Spins in One-Dimensional Optical Lattice}

\author{Xiong-Jun Liu}
\affiliation{Joint Quantum Institute, Department of Physics,
University of Maryland, College Park, Maryland 20742, USA}
\affiliation{Condensed Matter Theory Center, Department of Physics,
University of Maryland, College Park, Maryland 20742, USA}

\author{Zheng-Xin Liu}
\affiliation{Institute for Advanced Study, Tsinghua University, Beijing, 100084, P. R. China}
\affiliation{Department of Physics, Massachusetts Institute of Technology, Cambridge, Massachusetts 02139, USA}

\author{Meng Cheng}
\affiliation{Condensed Matter Theory Center, Department of Physics,
University of Maryland, College Park, Maryland 20742, USA}

\begin{abstract}
We propose to observe and manipulate topological edge spins in 1D optical lattice based on currently available experimental platforms.
Coupling the atomic spin states to a laser-induced periodic Zeeman field, the lattice system can be driven into a symmetry protected
topological (SPT) phase, which belongs to the chiral unitary (AIII) class protected by particle number conservation and chiral symmetries. In free-fermion case the SPT phase is classified by a $Z$ invariant which reduces to $Z_4$ with interactions.
The zero edge modes of the SPT phase are spin-polarized, with left and right edge spins polarized to opposite directions and forming
a topological spin-qubit (TSQ). We demonstrate a novel scheme to manipulate the zero modes and realize single spin control in optical lattice. The manipulation of TSQs has potential applications to quantum computation.
\end{abstract}
\date{\today }
\maketitle

\indent

{\it Introduction.}$-$Since the discovery of the
quantum Hall effect in two-dimensional (2D) electron gas \cite{QHE1}, the search for nontrivial topological states has become an exciting pursuit in condensed matter physics \cite{TP}. The recently observed time-reversal (TR) invariant topological insulators (TIs) have
opened a new chapter in the study of topological phases (TPs), attracting great efforts in both theory and experiments \cite{QSH1,QSH2}. Depending on whether the ground states have long-range or short-range entanglement,
the TPs can be classified into intrinsic or symmetry-protected topological (SPT) orders \cite{A-Z,Wen,Ryu}.
Being protected by the bulk gap, the intrinsic TPs are robust against any local perturbations, and the SPT phases are robust against those respecting given symmetries \cite{A-Z,Wen,Ryu,SPT}. This property may be applied to the fault-tolerant quantum computation \cite{TQC}.

While in theory there are numerous types of TPs, the existing topological orders in nature are rare. The recent great advancement in realizing effective spin-orbit (SO) interaction in cold atoms \cite{Liu,Lin,Chapman,Wang,MIT,Pan} opens intriguing new possibilities to probe SO effects \cite{SOC} and TPs in a controllable fashion. Theoretical proposals have been introduced in cold atoms for the study of TIs \cite{Liu1,Wu,Goldman,Li,Wuming}
and topological superfluids \cite{Chuanwei1,Sato,zhu1,Seo,Kraus}. Experimental studies of these exotic phases are, however, a delicate issue due to stringent conditions such as complicated lattice configurations or SO interactions. By far the only experimentally realized SO interaction \cite{Lin,Chapman,Wang,MIT,Pan} is the equal Rashba-Dresselhaus-type SO term as theoretically proposed by Liu etal \cite{Liu}. Therefore, how to observe nontrivial topological states with currently available experimental platforms is a central issue in the field of cold atoms \cite{Seo}.

In this letter, we propose to observe and manipulate topological edge spins in 1D optical lattice with SO interaction realizable in recent experiments \cite{Lin,Chapman,Wang,MIT}. The predicted SPT phase belongs to AIII class and is protected by $U(1)$ and chiral symmetries, with spin-polarized zero modes forming topological spin-qubits (TSQs). Our results may open the way to observe topological states of all ten Altand-Zirnbauer symmetry classes \cite{A-Z} with realistic cold atom systems, and have broad range of applications including realizing single spin control in optical lattice.

\begin{figure}[t]
\includegraphics[width=0.9\columnwidth]{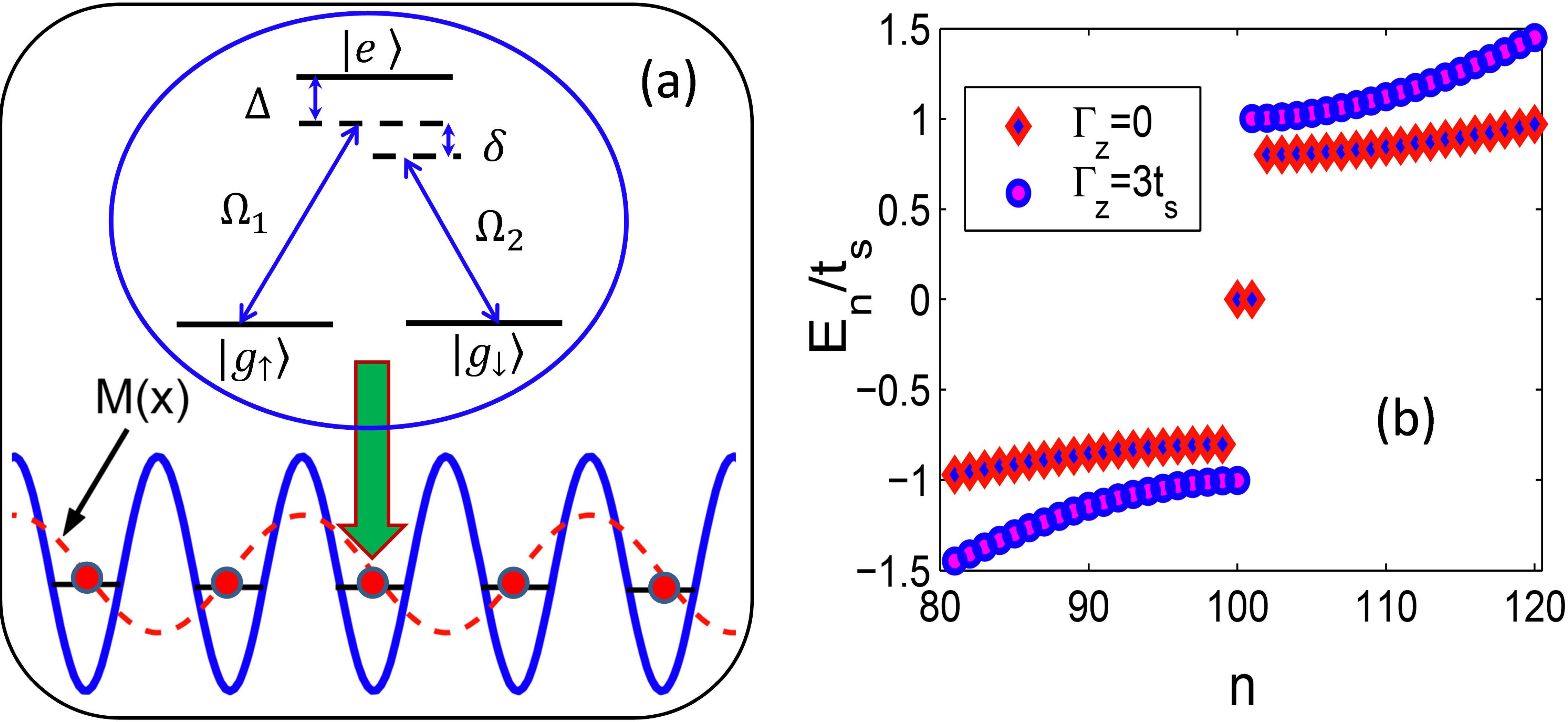} \caption{(Color online) (a) Cold fermions trapped in 1D optical lattice with internal three-level $\Lambda$-type configuration coupled to radiation. (b) Energy spectra with open boundary condition in the topological (diamond, $\Gamma_z=0$) and trivial (circle, $\Gamma_z=3t_s$) phases. The SO coupled hopping $t_{\rm so}^{(0)}=0.4t_s$.}
\label{lattice}
\end{figure}

{\it Model.}$-$Our model is based on quasi-1D cold fermions trapped in an optical lattice, with the internal three-level $\Lambda$-type configuration coupled to radiation, as shown in Fig.~\ref{lattice}. The transitions $|g_\uparrow\rangle,|g_\downarrow\rangle\rightarrow|e\rangle$ are driven by the laser fields with Rabi-frequencies $\Omega_1(x)=\Omega_{0}\sin(k_0x/2)$ and $\Omega_2(x)=\Omega_{0}\cos(k_0x/2)$, respectively. In the presence of a large one-photon detuning $|\Delta|\gg\Omega_{0}$ and a small two-photon detuning $|\delta|\ll\Omega_0$ for the transitions [Fig.~\ref{lattice}(a)], the Hamiltonian of the light-atom coupling system reads $H=H_0+H_1$, with $H_0=\sum_{\sigma=\uparrow,\downarrow}\bigr[\frac{p_x^2}{2m}+V_\sigma(x)\bigr]|g_\sigma\rangle\langle g_\sigma|+\hbar\delta|g_\downarrow\rangle\langle g_\downarrow|,
H_1=\hbar\Delta|e\rangle\langle e|-\hbar\bigr(\Omega_1|e\rangle\langle g_\uparrow|+\Omega_2|e\rangle\langle g_\downarrow|+{\rm H.c.}\bigr)$.
Here the diagonal potentials $V_{\uparrow,\downarrow}(x)$ are used to construct the 1D optical lattice, and $\sigma_{y,z}$ are the Pauli matrices in spin space. For $|\Delta|\gg\Omega_{0}$, the lasers $\Omega_{1,2}$ induce a two-photon Raman transition between $|g_{\uparrow}\rangle$ and $|g_\downarrow\rangle$. This configuration has been used to create the equal-Rashba-Dresselhaus SO interaction \cite{Liu,Lin,Chapman,Wang,MIT,Pan}.
The effect of the small two-photon detuning is equivalent to a Zeeman field along $z$ axis $\Gamma_z=\hbar\delta/2$, which in experiment can be precisely controlled with acoustic-optic modulator. Eliminating the excited state by $|e\rangle\approx\frac{1}{\Delta}(\Omega_1^*|g_\uparrow\rangle+\Omega_2^*|g_\downarrow\rangle)$ yields the effective Hamiltonian
\begin{eqnarray}\label{eqn:H2}
H_{\rm eff}&=&\frac{p_x^2}{2m}+\sum_{\sigma={1,2}}\bigr[V^{\rm Latt}_\sigma(x)+\Gamma_z\sigma_z\bigr]|g_\sigma\rangle\langle g_\sigma|-\nonumber\\
&&-\bigr[M(x)|g_\uparrow\rangle\langle g_\downarrow|+{\rm H.c.}\bigr],
\end{eqnarray}
where $M(x)=M_0\sin(k_0x)$ with $M_0=\frac{\hbar\Omega_0^2}{2\Delta}$ represents a transverse Zeeman field induced by the Raman process.

We next derive the tight-binding model. We consider first the $s$-band model in an optical lattice formed by the trapping potentials $V^{\rm Latt}_{\uparrow,\downarrow}(x)=-V_0\cos^2(k_0x)$, with the lattice trapping frequency $\omega=(2V_{0}k_0^2/m)^{1/2}$ \cite{RMPcoldatom}.
From the even-parity of the local $s$-orbitals $\phi_{s\sigma}$ ($\sigma=\uparrow,\downarrow$), the periodic term $M(x)$ does not couple the intrasite orbitals $\phi^{(i)}_{s\uparrow,\downarrow}$, but leads to a spin-flip hopping by $t_{\rm so}^{ij}=\int dx\phi^{(i)}_{s\uparrow}(x)M(x)\phi^{(j)}_{s\downarrow}(x)$ [see Fig.~\ref{lattice}(a)], representing the induced SO interaction. The spin-conserved hopping reads $t_s=\int dx\phi^{(j)}_{s\sigma}(x)\bigr[\frac{p_x^2}{2m}+V^{\rm}\bigr]\phi^{(j+1)}_{s\sigma}(x)$. Bearing these results in mind we write down the effective Hamiltonian in the tight-binding form: $H=-t_s\sum_{<i,j>,\sigma}\hat c_{i\sigma}^{\dag}\hat
c_{j\sigma}+\sum_{i}\Gamma_z(\hat n_{i\uparrow}-\hat n_{i\downarrow})+\bigr[\sum_{<i,j>}t_{\rm so}^{ij}\hat c_{i\uparrow}^\dag\hat c_{j\downarrow}+{\rm H.c.}]$,
with $\hat n_{i\sigma}=\hat c_{i\sigma}^\dag\hat c_{i\sigma}$. It can be verified that $t_{\rm so}^{j,j
\pm1}=\pm(-1)^jt_{\rm so}^{(0)}$, where $t_{\rm so}^{(0)}=\frac{\Omega_0^2}{\Delta}\int dx\phi_{s}(x)\sin(2k_0x)\phi_{s}(x-a)$ with $a$ the lattice constant.
Redefining the spin-down operator $\hat c_{j\downarrow}\rightarrow e^{i\pi x_j/a}\hat c_{j\downarrow}$, we recast the Hamiltonian into
\begin{eqnarray}\label{eqn:tightbinding2}
H&=&-t_s\sum_{<i,j>}(\hat c_{i\uparrow}^{\dag}\hat
c_{j\uparrow}-\hat c_{i\downarrow}^{\dag}\hat
c_{j\downarrow})+\sum_{i}\Gamma_z(\hat n_{i\uparrow}-\hat n_{i\downarrow})+\nonumber\\
&&+\bigr[\sum_{j}t_{\rm so}^{(0)}(\hat c_{j\uparrow}^\dag\hat c_{j+1\downarrow}-\hat c_{j\uparrow}^\dag\hat c_{j-1\downarrow})+{\rm H.c.}\bigr].
\end{eqnarray}
The above model can also be realized with $p$-band fermions in a different configuration of the optical lattice that $V^{\rm Latt}_{\uparrow,\downarrow}(x)=-V_0\sin^2(k_0x)$, which can be directly verified by noticing the odd-parity of $p$-orbitals. Remarkably, for $p$-band model the periodic Zeeman term $M(x)$ and the 1D lattice can be realized {\it simultaneously} by setting that $\Omega_1(x)=\Omega_{0}\sin(k_0x)$ and $\Omega_2=\Omega_{0}$ without applying additional lasers (see Supplementary Material \cite{SI} for details). This further greatly simplifies the experimental set-up and we believe that our proposal can be realized with realistic experimental platforms.

We analyze the symmetry of the Hamiltonian~\eqref{eqn:tightbinding2}. The TR and charge conjugation operators are respectively defined by ${\cal T}=iK\sigma_y$ with $K$ the complex conjugation, and ${\cal C}: (\hat c_{\sigma},\hat c^\dag_{\sigma})\longmapsto(\sigma_z)_{\sigma\sigma'}(\hat c^\dag_{\sigma'},\hat c_{\sigma'})$. One can check that while both ${\cal T}$ and ${\cal C}$ are broken in $H$, the chiral symmetry, defined as their product, is respected and $({\cal CT})H({\cal CT})^{-1}=H$, with $({\cal CT})^2=1$. Note the chiral symmetry is still preserved if a Zeeman term $\Gamma_y\sigma_y$ along $y$ axis is included in $H$. The complete symmetry group then reads $U(1)\times Z_2^T$, where $U(1)$ gives particle-number conservation and the anti-unitary group $Z_2^T$ is formed by $\{I,{\cal CT}\}$. The SPT phase of our free-fermion system belongs to the chiral unitary (AIII) class and is characterized by a $Z$ invariant \cite{A-Z,Wen,Ryu}. The $H$ can be rewritten in the $k$-space $H=-\sum_{k,\sigma\sigma'}\hat c_{k,\sigma}^{\dag}[d_z(k)\sigma_z+d_y(k)\sigma_y]_{\sigma,\sigma'}\hat
c_{k,\sigma'}$, with $d_y=2t_{\rm so}^{(0)}\sin(ka)$ and $d_z=-\Gamma_z+2t_s\cos(ka)$. This Hamiltonian describes a nontrivial topological insulator for $|\Gamma_z|<2t_s$ and otherwise a trivial insulator, with the bulk gap $E_g=\mbox{min}\{|2t_s-|\Gamma_z||, 2|t_{\rm so}^{(0)}|\}$ (Fig.~\ref{lattice} (b)).
In particular, when $\Gamma_{y,z}=0$ and $t_s=|t_{\rm so}^{(0)}|$, our model gives rise to a {\it flat band} with nontrivial topology.

{\it Edge states.}$-$The nontrivial topology can support degenerate boundary modes. Considering hard wall boundaries located at $x=0,L$, respectively \cite{hardwall} and diagonalizing $H$ in position space $H=\sum_{x_i}\mathcal{H}(x_i)$ with $\mathcal{H}(x_i)=-(t_s\sigma_z+it^{(0)}_{\rm so}\sigma_y)\hat c_{x_i}^\dag\hat
c_{x_i+a}+\Gamma_z\sigma_z\hat c_{x_i}^\dag\hat
c_{x_i}+h.c.$, we obtain the edge state localized on left
boundary $x=0$ as
\begin{eqnarray}\label{eqn:gapstate1}
\psi_L(x_i)=\frac{1}{\sqrt{{\cal N}}}[(\lambda_+)^{x_i/a}-(\lambda_-)^{x_i/a}]|\chi_+\rangle,
\end{eqnarray}
and accordingly the one on $x=L$ by $\psi_R(x_i)=\frac{1}{\sqrt{{\cal N}}}[(\lambda_+)^{(L-x_i)/a}-(\lambda_-)^{(L-x_i)/a}]|\chi_-\rangle$. Here ${\cal N}$ is the normalization factor, the spin eigensates $\sigma_x|\chi_\pm\rangle=\pm|\chi_\pm\rangle$, and $\lambda_\pm=(\Gamma_z\pm\sqrt{\Gamma_z^2-4t_s^2+4|t^{(0)}_{\rm so}|^2})/(2t_s+2|t^{(0)}_{\rm so}|)$. Therefore the two edge modes are polarized to the opposite $\pm x$ directions. Note $\psi_{L}$ and $\psi_{R}$ span the complete Hilbert space of {\it one single} $1/2$-spin or spin-qubit. Each edge state equals {\it one-half} of a single spin, similar to the relation between a Majorana fermion and a complex fermion in topological superconductors. As a result, we expect the robustness of the zero modes to any local operations without breaking the $U(1)$ and ${\cal CT}$ symmetries \cite{note0}. These properties of the TSQ may be applicable to fault-tolerant quantum computation \cite{TQC}. Moreover, the $Z$ classification implies that single-particle couplings respecting $U(1)$ and ${\cal CT}$ cannot gap out the edge modes in arbitrary $N$-chain system of 1D lattices. Interestingly, however, we have confirmed that with weak interactions a system with up to four chains of the 1D lattices can be adiabatically connected to a trivial phase without closing the bulk gap, implying that the $Z$ classification breaks down to $Z_4$ with interactions \cite{SI}. This result suggests an interesting platform to study the classification of SPT phases with cold atoms.
\begin{figure}[ht]
\includegraphics[width=0.9\columnwidth]{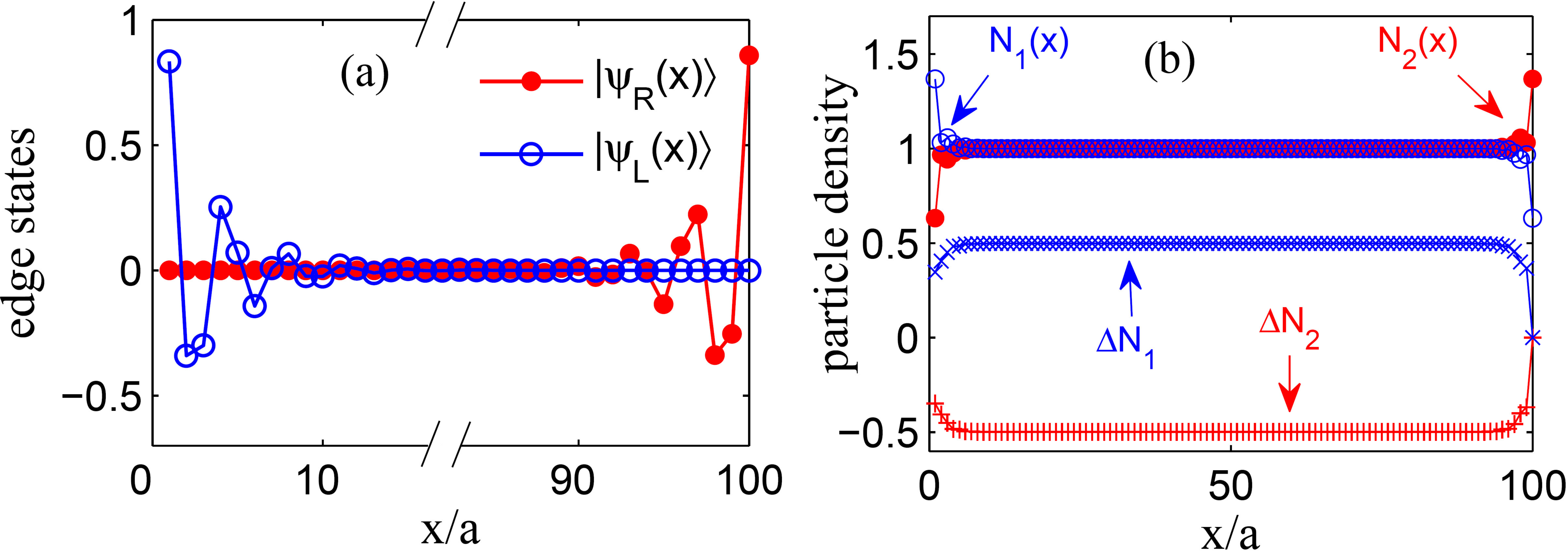} \caption{(Color online) (a) Wave functions for zero modes $|\psi_{L,R}\rangle$; (b) $1/2$-particle fractionalization (seen by $\Delta N_{1,2}$) for zero modes. The parameters $t_{\rm so}^{(0)}=0.4t_s$ and $\Gamma_z=0.3t_s$, with which the localization length of bound modes $\xi_0$=2.36a.}
\label{zeromodes}
\end{figure}

Existence of zero modes leads to particle fractionalization, which is another direct observable in experiment. A zero mode is contributed half from the valence band and half from the conduction band. Therefore, an edge state carries $+1/2$ ($-1/2$) particle if it is occupied (unoccupied) \cite{SI}. This result is confirmed by numerical simulation shown in Fig.~\ref{zeromodes}, where we calculate $\Delta N_{1,2}(x_i)=\int^{x_i}_0dx'N_{1,2}(x')-x_i/a$ at half-filling, with $N_{1(2)}(x)$ the density of fermions when the left (right) edge mode is filled. The fermion number carried by an occupied (unoccupied) edge mode is then given by $n_{1(2)}=\Delta N_{1(2)}(\xi)$ with $\xi\gg\xi_0$. Here $\xi_0=-a/\ln|\lambda_+|$ is the localization length of $\psi_{L,R}$. The $1/2$-fractionalization is clearly seen when $\xi$ is several times greater than $\xi_0$ (Fig.~\ref{zeromodes} (b)). Being a topological invariant, the $1/2$-fractionalization can be confirmed to be robust against weak disorder scatterings without breaking the given symmetries.

%

{\it Correlation effects.}$-$A particular advantage in cold atoms is that one can investigate correlation effects on the predicted SPT phase by precisely controlling the interaction. For spin-$1/2$ cold fermions the onsite Hubbard interaction $U\sum_i n_{i\uparrow}n_{i\downarrow}$ can be well controlled by Feshbach resonance \cite{RMPcoldatom}. For a single-chain system, we expect that the topological phase is stable against weak interactions relative to the single-particle bulk gap, while the strong repulsive interaction can always drive the system into a Mott insulating phase. The correlation effects around the critical point can be probed by Abelian bosonization approach combined with renormalization group (RG) analysis \cite{book}. Note in the non-interacting regime, the phase diagram of the single-chain system is determined by the SO and Zeeman terms which define two mass terms $\mathcal{H}_\text{SO}=\frac{u}{\pi a}\sin\sqrt{2}\phi_\rho\cos\sqrt{2}\theta_\sigma, \mathcal{H}_\text{Z}=\frac{w}{\pi a}\cos\sqrt{2}\phi_\rho\sin\sqrt{2}\theta_\sigma$ in the bosonized Hamiltonian, where the masses $u=2t^{(0)}_\text{so}$, $w=\Gamma_y$, and $\phi_{\rho,\sigma},\theta_{\rho,\sigma}$ are boson representation of the fermion fields \cite{SI}. The fate of the system in the presence of the interaction depends on which mass term flows first to the strong coupling regime under RG.

A direct power counting shows the same RG flow for the masses $u$ and $w$ in the first-order perturbation. Therefore the next-order perturbation expansion is necessary to capture correctly the fate of the topological phase transition. By deriving the RG flow equations up to one-loop order~\cite{RG}, we find the renormalization to $u, w$, the umklapp scattering $g_\rho$ and spin backscattering $g_\sigma$ by \cite{SI}:
\begin{equation}
  \begin{split}
	\frac{\mathrm{d}u}{\mathrm{d} l}&=\frac{3-K_\rho}{2}u-\frac{g_\rho u}{4\pi v_F}+\frac{g_\sigma u}{4\pi v_F},\\
	\frac{\mathrm{d}w}{\mathrm{d} l}&=\frac{3-K_\rho}{2}w +\frac{g_\rho w}{4\pi v_F}+\frac{g_\sigma w}{4\pi v_F},\\
	\frac{\mathrm{d}g_\rho}{\mathrm{d} l}&=\frac{g_\rho^2}{\pi v_F}, \ \ \
	\frac{\mathrm{d}g_\sigma}{\mathrm{d} l}=\frac{g_\sigma^2}{\pi v_F},
\end{split}
  \label{}
\end{equation}
where the bare values of the coupling constants $g_\rho=-g_\sigma=U, u=2t^{(0)}_\text{so}, w=\Gamma_y$,
and $l$ is the logarithm of the length scale. 
The renormalization of Luttinger parameter $K_\rho$ has been neglected as it is a higher order correction.
For $U>0$, $g_\sigma$ marginally flows to zero so we drop it off below. This is consistent with the result that repulsive interaction cannot gap out the spin sector in the 1D Hubbard model. $g_\rho$ is marginally relevant and can be solved by $g_\rho(l)=\frac{\pi v_F g_\rho(0)}{\pi v_F-g_\rho(0)l}$.
Substituting this result into RG equations of $u$ and $w$ yields after integration $u(l)=u(0)[1-\frac{g_\rho(0)l}{\pi v_F}]^{\frac{1}{4}}e^{(3-K_\rho)l/2},
w(l)=w(0)[1+\frac{g_\rho(0)l}{\pi v_F}]^{\frac{1}{4}}e^{(3-K_\rho)l/2}$. The physics is clear: the repulsive interaction ($g_\rho>0$) suppresses SO induced mass term $u$ while enhances the trivial mass term $w$.
%
%
The fate of the system depends on which of $u$ and $w$ reaches the strong-coupling regime first. Assuming $|g_\rho(0)l|\ll v_F$, we find the TP transition occurs at
\begin{equation}
  u(0)=\bigr[w(0)\bigr]^\gamma, \: \gamma\approx 1-\frac{g_\rho(0)}{4\pi v_F(3-K_\rho)}.
  \label{}
\end{equation}
This gives the scaling law at the phase boundary with interaction.
Note $\gamma<1$ for $U>0$. The above scaling relation implies that a repulsive interaction suppresses the SPT phase. Accordingly, if initially the noninteracting system is topologically nontrivial with $u(0)>w(0)>0$, increasing $U$ to the regime $u(0)<[w(0)]^\gamma$ drives the system into a trivial phase.

\begin{figure}[ht]
\includegraphics[width=0.9\columnwidth]{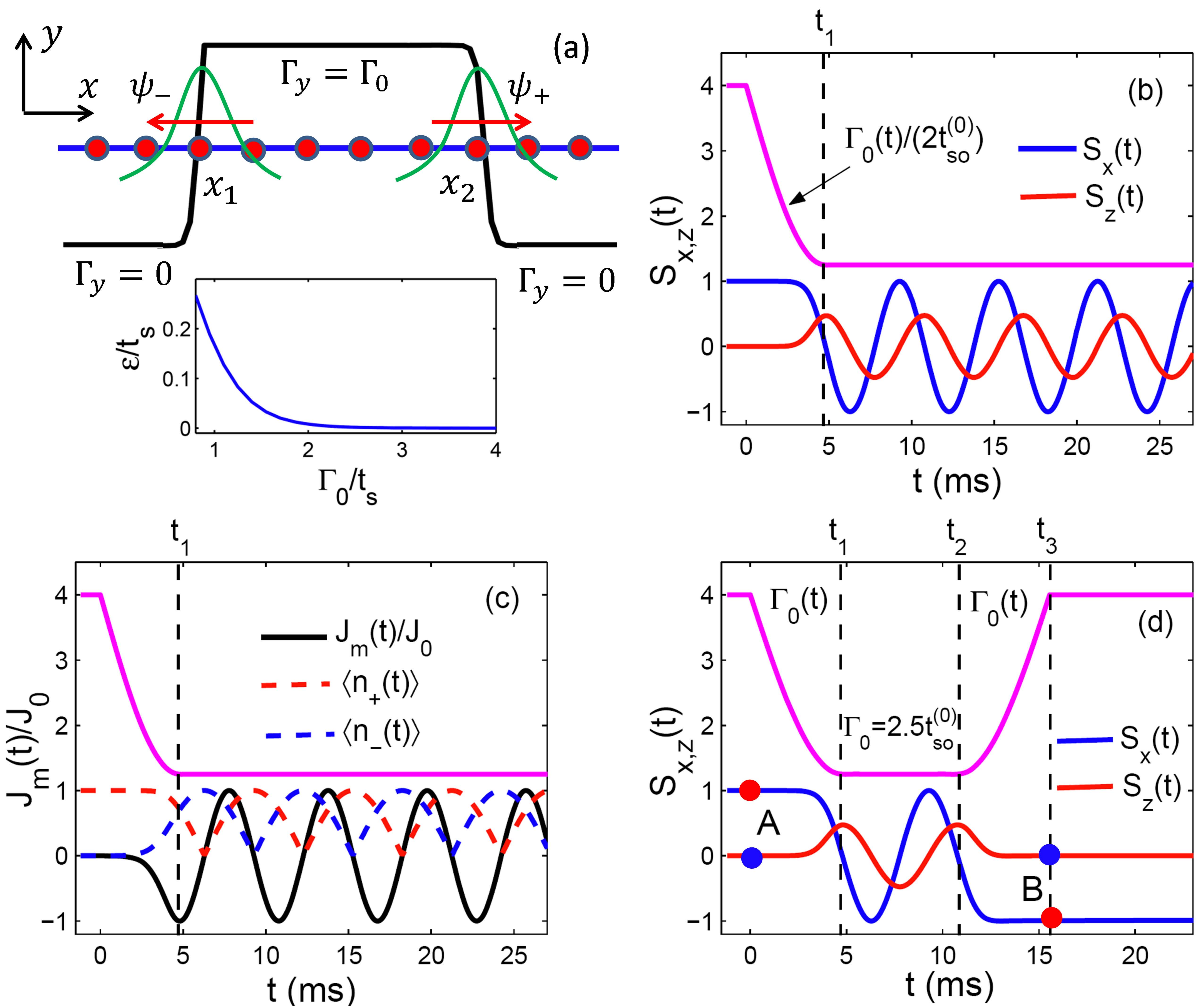} \caption{(Color online) Spin Rabi-oscillations with the parameters $t_s=3.15$kHz, $t_{\rm so}^{(0)}=0.4t_s$, and $\Delta x=10a\sim4\mu{m}$. (a) Mass domain created by setting $|\Gamma_0|>2t_{\rm so}^{(0)}$ for $x_1<x<x_2$ which localizes a spin-qubit composed of two bound modes $|\psi_\pm\rangle$ on $x=x_1,x_2$, respectively; (b) Spin Rabi-oscillation by smoothly reducing $|\Gamma_0|$ from $8t_{\rm so}^{(0)}$ to $2.5t_{\rm so}^{(0)}$; (c) The mass current $J_m(t)$ and expectation values of particle numbers $\langle n_\pm(t)\rangle$ in states $|\psi_\pm\rangle$; (d) Spin-flip operation by controlling that $\gamma(t_3)=(2m+1)\pi$ with $m=1$. The initial spin state $|\chi_+\rangle$ (Points A) flips to be $|\chi_-\rangle$ (Points B).}
\label{oscillation}
\end{figure}

{\it Single spin control.}$-$Now we study an interesting application of the present results to realizing single spin control. Besides the edge modes localized on the ends, TSQs can also be obtained in the middle areas by creating mass domains in the lattice. This can be achieved by applying a local Zeeman term $\Gamma_{y}$ or $\Gamma_z$. For example, we consider $\Gamma_z=0$ everywhere, but $\Gamma_y=\Gamma_0$ for $x_1<x<x_2$ and $\Gamma_y=0$ otherwise. The local $\Gamma_y$ can be generated by applying another two lasers which cross with the 1D lattice and couple the atoms in the area $x_1<x<x_2$ to induce a local resonant Raman coupling between $|g_{\uparrow}\rangle$ and $|g_\downarrow\rangle$ (Fig.~\ref{oscillation}(a)). Employing a $\pi/2$-phase offset in the Rabi-frequencies of the two lasers, the Raman coupling takes the form $\Gamma_0\sigma_y$, with $\Gamma_0$ controlled by the laser strength. When $|\Gamma_0|>2|t_{\rm so}^{(0)}|$ a mass domain is created, associated with two midgap spin states $|\psi_\pm\rangle$ respectively localized around $x=x_{1,2}$ (Fig.~\ref{oscillation}(a)). The width $\Delta x=x_2-x_1$ and height of the domain are respectively adjusted by the waist size and strength of the two laser beams. Due to the nonlocality of the TSQ, creation of a single qubit here is not restricted by the size of the laser beams. This is a fundamental difference from creating conventional single qubit by optical dipole trapping which requires tiny-sized laser beams to reach a very small trapping volume \cite{Schlosser}. Note in realistic case the laser induced $\Gamma_y$ may vary fast but not in the form of step functions around $x=x_{1,2}$, which, however, does not affect the main results presented here \cite{SI}. Coupling between $|\psi_+\rangle$ and $|\psi_-\rangle$ results in an energy splitting $2{\cal E}\propto e^{-(|\Gamma_0|-2|t_{\rm so}^{(0)}|)\Delta x/(2at_s)}$, which is controlled by $\Gamma_0$ and $\Delta x$ (Fig.~\ref{oscillation}(a), lower panel). In the limit $(|\Gamma_0|-2|t_{\rm so}^{(0)}|)\Delta x/(2at_s)\gg1$, such coupling is negligible, and the two zero modes consist of a single spin qubit which is topologically stable. Let $|\psi_+\rangle$ be initially occupied while $|\psi_-\rangle$ be left vacancy. Reducing $|\Gamma_0|$ smoothly can open the coupling in $|\psi_\pm\rangle$ and lead to spin state evolving as \cite{note} $|\psi(t)\rangle=\alpha(t)\varphi_-(x-x_1)|\chi_-\rangle+\beta(t)\varphi_+(x-x_2)|\chi_+\rangle$, with $\alpha(0)=0, \beta(0)=1$, and $\varphi_\pm$ the spatial parts of the bound state wave-functions. The spin-polarization densities are given by $s_{x,y,z}(x,t)=\langle\psi(x)|\sigma_{x,y,z}|\psi(x)\rangle$, and the spin expectation values $S_{x,y,z}(t)=\int dxs_{x,y,z}(x,t)$. It can be verified that $S_y(t)=0$, and
\begin{equation}
  \begin{split}
S_{x}(t)&=|\alpha(t)|^2-|\beta(t)|^2,\\
S_z(t)&=2\mbox{Re}\bigr[\alpha(t)\beta^*(t)\int dx\varphi^*_+(x)\varphi_-(x)\bigr].
\end{split}
  \label{}
\end{equation}
This phenomenon is analogous to spin precession with the rotating angle yielding $\gamma(t)=2\int_0^tdt'{\cal E}(t')$. We have then $\alpha=\cos\gamma(t)$ and $\beta=\sin\gamma(t)$. The amplitude of $S_{z}(t)$ is given by $S_{z}^{\rm max}=|\int dx\varphi^*_+(x)\varphi_-(x)|$, which measures the overlapping integral of $\varphi_\pm$. Accordingly, if we apply the local Zeeman field along $z$ rather than $y$ axis, we shall obtain that the spin evolves in the $x$-$y$ plane. Note the spin Rabi-oscillation is induced by quantum tunneling. Therefore it is associated with a tunneling current given by $J_m(t)=-\frac{\Delta x{\cal E}}{2\pi\hbar}\partial_t|\alpha(t)|^2$ between $x_1$ and $x_2$. In experiment the internal states of a single atom can be detected without energy transfer \cite{Volz}, which is applicable to observe the spin Rabi-oscillations, while the oscillation of $S_x(t)$ can be more conveniently observed by measuring the number of fermions $\langle n_\pm(t)\rangle$ localized around $x_{1,2}$ with single-site resolution technology \cite{SSR}, and $J_m(t)$ can be detected by measuring the change rate with time of such fermion numbers.

We show in Fig.~\ref{oscillation} (b-d) the numerical simulation for single spin control with the parameter regime that $t_s=3.15$kHz, $t_{\rm so}^{(0)}=0.4t_s$, and $\Delta x=10a$. For $t<0$, $\Gamma_0=8t_{\rm so}^{(0)}$ and the coupling in $|\psi_\pm\rangle$ is negligible. Reducing $\Gamma_0$ at $t>0$ leads to spin evolution and by fixing $\Gamma_0=2.5t_{\rm so}^{(0)}$ for $t>t_1$ the spin oscillates with a period of $5.984$ms (b-c). Note the quantum state of the spin can be precisely controlled by properly manipulating $\gamma(t)$.
For example, in Fig.~\ref{oscillation} (d) we demonstrate the spin-flip operation $|\chi_+\rangle\rightarrow|\chi_-\rangle$ by requiring $\gamma(t_3)=(2m+1)\pi$. Here $m\in Z$ and in (d) we take $m=1$.
Note one may integrate multiple TSQs with e.g. atom-chip technology and individually control them by creating multiple mass domains in the 1D lattice. The precise manipulation of such integrated TSQs may have interesting applications in developing scalable spin-based quantum computers.

Before conclusion we estimate the parameter values for realistic experimental observations. For example, in $^{40}$K atoms we have the recoil energy $E_R/\hbar=\hbar k_0^2/2m=48$kHz using red-detuned lasers of wavelength $773$nm to form the optical lattice. Taking that $V_0=5E_R$ and $M_0=2E_R$, we have that the lattice trapping frequency $\omega=214$kHz, and hopping coefficients $t_s/\hbar\simeq3.15$kHz and $t_{\rm so}^{(0)}/\hbar\simeq1.3$kHz. Then the bulk gap equals $E_g/\hbar=2.6$kHz for $\Gamma_z=0$, indicating a temperature $T=19$nK for the experimental observation. Also, under this parameter regime the life time of the atoms is over $1.0$s, which is long enough for the detection and manipulation of the topological edge spins.

{\it Conclusions.}$-$We have proposed to observe and manipulate SPT phase of AIII class in 1D optical lattice, and demonstrated single spin control by manipulating spin-polarized zero modes which is applicable to spin-based quantum computation. The minimum requirement for the proposed scheme is a regular 1D lattice and a transverse Zeeman field, which can be realized simultaneously in a single two-photon Raman transition as used in the recent experiments \cite{Lin,Chapman,Wang,MIT,Pan}. The present study may open the search for topological states of all ten Altand-Zirnbauer symmetry classes with realistic cold atom systems, and its remarkable feasibility will attract both theoretical and experimental efforts in future.

We thank Tin-Lun Ho, X. G. Wen, X. Chen, V. W. Liu, and Cenke Xu for helpful discussions. We acknowledge the support from  JQI-NSF-PFC, Microsoft-Q, and DARPA-QuEST.


\noindent

\newpage
\begin{widetext}

\begin{appendix}

\section*{Manipulating Topological Edge Spins in One-Dimensional Optical Lattice --- Supplementary Material}

In this supplementary material we provide the details of some results in the main text.

\section{Tight-binding Hamiltonian for $p$-band model}

\begin{figure}[b]
\includegraphics[width=0.45\columnwidth]{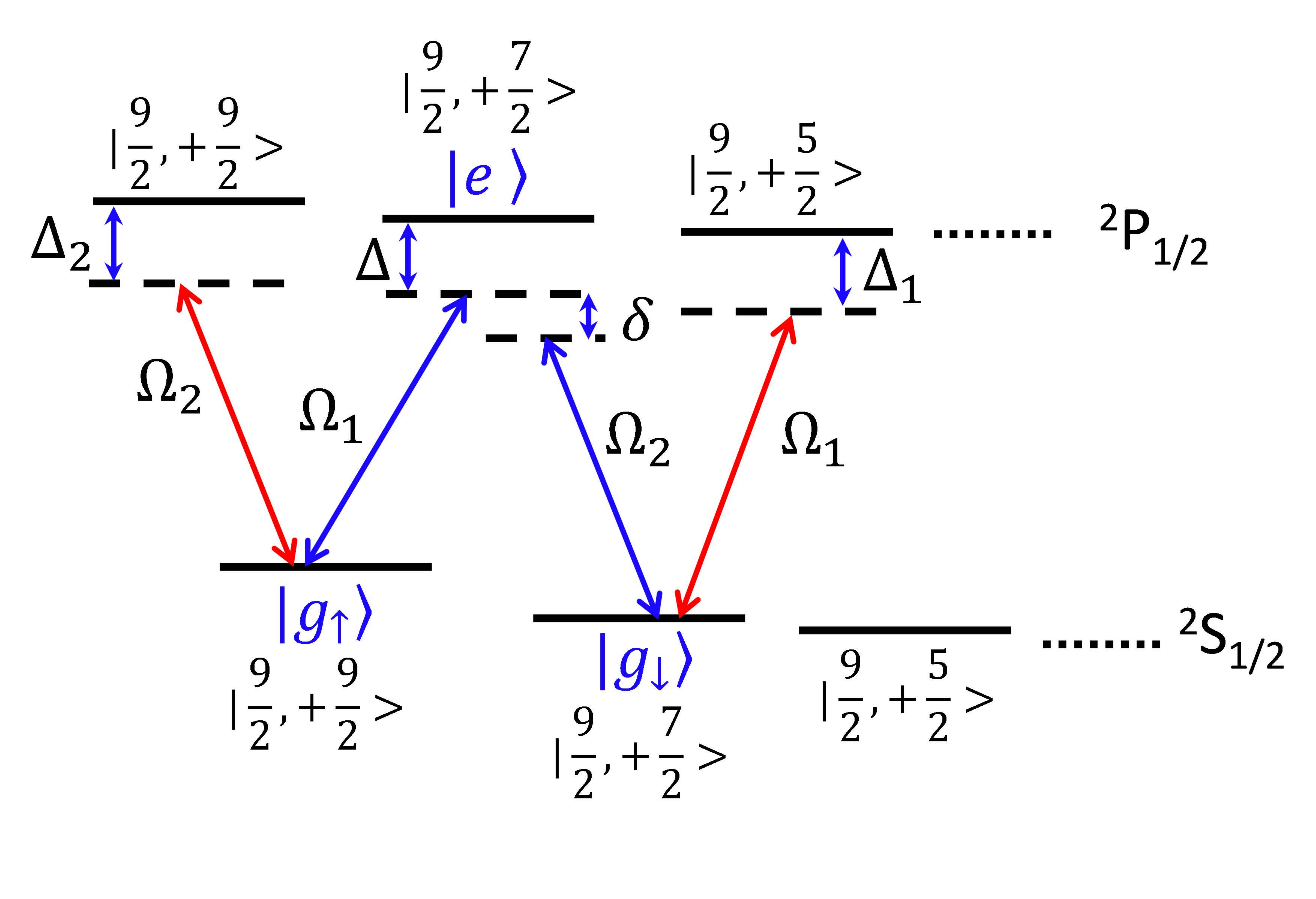} \caption{(Color online) Laser induced transitions in the realistic $^{40}$K atoms.}
\label{p-band}
\end{figure}
In this section we provide details of deriving the tight-binding Hamiltonian for the $p$-band model. As mentioned in the main text, only two lasers $\Omega_{1,2}$, with $\Omega_1(x)=\Omega_0\sin(k_0x)$ and $\Omega_2=\Omega_0$ used to induce the two-photon Raman transition, are needed to generate simultaneously the periodic transverse Zeeman term and the 1D optical lattice. This further greatly simplifies the set-up for the experimental realization. Fig.~\ref{p-band} shows the realistic transitions for $^{40}$K atoms induced by $\Omega_{1,2}$. It is noteworthy that besides the induced $\Lambda$-configuration, $\Omega_1$ and $\Omega_2$ can also couple respectively to the states $|g_\downarrow\rangle$ and $|g_\uparrow\rangle$ (i.e. $|\frac{9}{2},+\frac{7}{2}\rangle$ and $|\frac{9}{2},+\frac{9}{2}\rangle$) (Fig.~\ref{p-band}), while these couplings cannot lead to additional Raman transitions between the ground (pseudo-)spin-$1/2$ subspace and other ground states since such transitions are associated with large two-photon detunings (see the experiments Ref. [11-15] in the main text. The typical two-photon detuning for such processes is $10$MHz, much larger than the Raman transition strength which is about $0.1$MHz). Therefore the two couplings only lead to additional diagonal optical potentials $V_\uparrow=-|\Omega_2|^2/\Delta_2$ and $V_\downarrow=-|\Omega_1|^2/\Delta_1$ for $|g_\uparrow\rangle$ and $|g_\downarrow\rangle$, respectively. The light-atom coupling Hamiltonian for the realistic $^{40}$K system then reads
\begin{eqnarray}\label{eqn:H1-SI}
H&=&\sum_{\sigma=\uparrow,\downarrow}\bigr[\frac{p_x^2}{2m}+V_\sigma(x)\bigr]|g_\sigma\rangle\langle g_\sigma|+\hbar\delta|g_\downarrow\rangle\langle g_\downarrow|+\nonumber\\
&&+\hbar\Delta|e\rangle\langle e|-\hbar\bigr(\Omega_1|e\rangle\langle g_\uparrow|+\Omega_2|e\rangle\langle g_\downarrow|+{\rm H.c.}\bigr).
\end{eqnarray}
For the large one-photon detuning condition $|\Delta|\gg|\Omega_0|$, we can eliminate the excited state by $|e\rangle\approx\frac{1}{\Delta}(\Omega_1^*|g_\uparrow\rangle+\Omega_2^*|g_\downarrow\rangle)$ and obtain
\begin{eqnarray}\label{eqn:H2-SI}
H_{\rm eff}&=&\frac{p_x^2}{2m}+\sum_{\sigma={1,2}}\bigr[V^{\rm Latt}_\sigma(x)+\Gamma_z\sigma_z\bigr]|g_\sigma\rangle\langle g_\sigma|-\nonumber\\
&&-\bigr[M_0\sin(k_0x)|g_\uparrow\rangle\langle g_\downarrow|+{\rm H.c.}\bigr],
\end{eqnarray}
with $M_0=\hbar\Omega_0^2/\Delta$ and
\begin{eqnarray}\label{eqn:H2-SI}
V^{\rm Latt}_\uparrow(x)=-\hbar\frac{\Omega_0^2}{\Delta_2}-\hbar\frac{\Omega_0^2}{\Delta}\sin^2(k_0x), \ \ V^{\rm Latt}_\downarrow(x)=-\hbar\frac{\Omega_0^2}{\Delta}-\hbar\frac{\Omega_0^2}{\Delta_1}\sin^2(k_0x).
\end{eqnarray}
Note in realistic experiments the difference between $\Delta_{1,2}$ and $\Delta$ (about $10$MHz) is negligible relative to their magnitudes (in the order of $10^4$GHz). We can take that $\Delta_1=\Delta_2=\Delta$, and then get (neglecting the constant terms)
\begin{eqnarray}\label{eqn:H2-SI}
V^{\rm Latt}_\uparrow(x)=V^{\rm Latt}_\downarrow(x)=-\hbar\frac{\Omega_0^2}{\Delta}\sin^2(k_0x).
\end{eqnarray}
The tight-binding Hamiltonian for $p$-band model can be derived straightforwardly. 
By noticing the odd-parity of the local $p$-orbitals, the periodic term $M(x)$ leads to the spin-flip hopping by
$t_{\rm so}^{ij}=\int dx\phi^{(i)}_{p\uparrow}(x)M(x)\phi^{(j)}_{p\downarrow}(x)=\pm(-1)^jt_{\rm so}^{(0)}$, with $t_{\rm so}^{(0)}=\frac{\Omega_0^2}{\Delta}\int dx\phi_{p}(x)\sin(k_0x+\frac{\pi}{2})\phi_{p}(x-a)$. The spin-conserved hopping reads $t_s=\int dx\phi^{(j)}_{p\sigma}(x)\bigr[\frac{p_x^2}{2m}+V^{\rm}\bigr]\phi^{(j+1)}_{p\sigma}(x)$. With these results we can finally get the tight-binding Hamiltonian
\begin{eqnarray}\label{eqn:tightbinding2-SI}
H&=&-t_s\sum_{<i,j>}(\hat c_{i\uparrow}^{\dag}\hat
c_{j\uparrow}-\hat c_{i\downarrow}^{\dag}\hat
c_{j\downarrow})+\sum_{i}\Gamma_z(\hat n_{i\uparrow}-\hat n_{i\downarrow})+\nonumber\\
&&+\bigr[\sum_{j}t_{\rm so}^{(0)}(\hat c_{j\uparrow}^\dag\hat c_{j+1\downarrow}-\hat c_{j\uparrow}^\dag\hat c_{j-1\downarrow})+{\rm H.c.}\bigr],
\end{eqnarray}
which is the same as Eq.~(2) in the main text. More generally, with this configuration all the bands with odd-parity local orbitals $(p,f,...)$ can be described by the tight-binding Hamiltonian~\eqref{eqn:tightbinding2-SI}, while those bands with even-parity local orbitals ($s,d,...$) are always topologically trivial. Accordingly, for another configuration with the lattice potentials $V^{\rm Latt}_{\uparrow,\downarrow}(x)=-V_0\cos^2(k_0x)$ considered in the main text, all the bands with even-parity local orbitals can be described by the above Hamiltonian~\eqref{eqn:tightbinding2-SI}, while the other bands with odd-parity local orbitals are always topologically trivial.

\section{Topological Classification}

\subsection{Noninteracting regime}

With the inclusion of both $\Gamma_y$ and $\Gamma_z$, the generic Hamiltonian obtained in the main text is given by
\begin{eqnarray}\label{eqn:aptightbinding}
H&=&-\sum_{j}\bigr[t_s(\hat c_{j\uparrow}^{\dag}\hat
c_{j+1\uparrow}-\hat c_{j\downarrow}^{\dag}\hat
c_{j+1\downarrow})-t_{\rm so}^{(0)}(\hat c_{j\uparrow}^\dag\hat c_{j+1\downarrow}-\hat c_{j\uparrow}^\dag\hat c_{j-1\downarrow})+{\rm H.c.}\bigr]+\sum_{j}\bigr[\Gamma_z(\hat n_{j\uparrow}-\hat n_{j\downarrow})-\Gamma_y(i\hat c^\dag_{j\uparrow}c_{j\downarrow}+{\rm H.c.})\bigr]\nonumber\\
&=&\sum_{k,\sigma}\hat c_{k,\sigma\sigma'}^{\dag}{\cal H}_{\sigma\sigma'}(k)\hat
c_{k,\sigma'},
\end{eqnarray}
where ${\cal H}(k)=d_y(k)\sigma_y+d_z(k)\sigma_z$ with $d_y=\Gamma_y+2t_{\rm so}^{(0)}\sin(ka)$ and $d_z=\Gamma_z-2t_s\cos(ka)$.
The full symmetry of the system is $U(1)\times (CT)$, where the phase transformation operator $U(1)$, charge conjugation operator $C$ and time reversal operator $T$ are defined as:
\begin{eqnarray*}
&&U(\theta)\hat c U(\theta)^{-1}= e^{i\theta}\hat c, \ \ \ \ \ {\cal C}\hat c{\cal C}^{-1}=\sigma_z\hat c^\dag;\\
&&{\cal T}\hat c{\cal T}^{-1}=-i\sigma_y K \hat c, \ \ \ ({\cal CT})\hat c({\cal CT})^{-1}=\sigma_x \hat c^\dag.
\end{eqnarray*}
The following commutation relations can be checked: ${\cal C}^2=-{\cal T}^2=({\cal CT})^2=1, \{{\cal C},{\cal T}\}=0$, and $[{\cal CT}, U(1)]=0$.
The subgroup $\{I, {\cal CT}\}$ is anti-unitary and can be denoted as $Z_2^T$, and the complete symmetry group can also be written as $U(1)\times Z_2^T$.
Owning to this symmetry group, the free fermion system belongs to chiral unitary (AIII) class and is characterized by a $Z$ invariant in the noninteracting case.

If the Zeeman field along $y$ axis vanishes, i.e. $\Gamma_y=0$, alternatively the symmetry group can be chosen as $U(1)\rtimes(C\times T)$ where $U(\theta)\hat c U(\theta)^{-1}= e^{i\theta}\hat c$, $\mathcal{C}\hat{c}{\mathcal{C}}^{-1}=\sigma_x {\hat{c}}^\dag$, ${\cal T}\hat c{\cal T}^{-1}=K \hat{c}$, with ${\cal C}^2={\cal T}^2=({\cal CT})^2=1$, and $[\mathcal C, \mathcal T]=0, \mathcal T U(\theta)=U(-\theta)\mathcal T, \mathcal C U(\theta)=U(-\theta)\mathcal C$.
In this case both $T$ and $C$ are symmetries of the Hamiltonian and the system then belongs to the BDI class which is also classified by a $Z$ invariant in the non-interacting case.

\subsection{Interacting regime}

For a N-chain system under the half-filling condition, the total ground-state degeneracy is $C_{2N}^N$ without interactions. Let $\hat c_{m,L/R}$ and $\hat c^\dag_{m,L/R}$ be the annihilation and creation operators of the left/right edge mode for the $m$-th chain, respectively, where $\hat c_L=(\hat c_\uparrow+\hat c_\downarrow)/\sqrt2$ and $\hat c_R=(\hat c_\uparrow-\hat c_\downarrow)/\sqrt2$. The edge states of the $m$-th chain can be written as $\hat c_{m,L}^\dag |0\rangle_{m}$ and $c_{m,R}^\dag |0\rangle_m$, where $|0\rangle_m$ is the ground state for the bulk.

An interesting question is what happens if we turn on interactions. It turns out that in the presence of interactions, the $Z$ classification breaks down to $Z_4$. To confirm this result, we will study the ground states of a N-chain system step by step.

First, for $N=2$, a generic $U(1)\times (CT)$-symmetric interaction between the edge zero modes reads
\begin{eqnarray}\label{2chain}
H_{12}&=&\sum_{s=L,R}\left[V_{12}^{(4)}\hat c_{1,s}^\dag \hat c_{1,s}\hat c_{2,s}^\dag \hat c_{2,s}+ (V_{12}^{(4)})^*\hat c_{1,s} \hat c^\dag_{1,s}\hat c_{2,s} \hat c^\dag_{2,s}\right]\nonumber\\
&=&\sum_{s}V_{12}^{(4)}(2\hat c_{1,s}^\dag \hat c_{1,s}\hat c_{2,s}^\dag \hat c_{2,s}-\hat c^\dag_{1,s}\hat c_{1,s}-\hat c^\dag_{2,s} \hat c_{2,s}+1).
\end{eqnarray}
With above interaction the minimum degeneracy of the 2-chain system is two-fold, which is obtained when $V_{12}^{(4)}=(V_{12}^{(4)})^*=-V_0<0$. It is straightforward to check that the states $|01\rangle_{1}|01\rangle_{2}$ and $|10\rangle_{1}|10\rangle_{2}$ have energy $-2V_0$ while the states $|10\rangle_{1}|01\rangle_{2}$, $|01\rangle_{1}|10\rangle_{2}$,$|00\rangle_{1}|11\rangle_{2}$, and $|11\rangle_{1}|00\rangle_{2}$ have a higher energy $0$. Then the ground state is two-fold degenerate, as given by $|01\rangle_{1}|01\rangle_{2}$ and $|10\rangle_{1}|10\rangle_{2}$.

Second, for $N=3$, the ground state degeneracy $C_6^3=20$ can be reduced to 2 by two-chain interactions according to (\ref{2chain}): $H_{\rm int}=H_{12}+H_{23}$. For instance, if $V_{12}^{(4)}=V_{23}^{(4)}=-V_0<0$, the two-fold degenerate ground states are
\begin{eqnarray}\label{GS3_1}
|01\rangle_1|01\rangle_2|01\rangle_3,\ \  |10\rangle_1|10\rangle_2|10\rangle_3;
 \end{eqnarray}
and if $V_{12}=-V_{23}=-V_0<0$, the ground states read
\begin{eqnarray}\label{GS3_2}
|01\rangle_1|01\rangle_2|10\rangle_3,\ \  |10\rangle_1|10\rangle_2|01\rangle_3.
\end{eqnarray}

However, for the 3-chain system, one should also consider three-body interactions. One of the possible 3-chain interactions reads
\begin{eqnarray}
H_{123}=\sum_{s}(V_{123}^{(6)}\hat c_{1,s}^\dag \hat c_{1,s}\hat c_{2,s}^\dag \hat c_{2,s}\hat c_{3,s}^\dag \hat c_{3,s}+V_{123}^{(6)}\hat c_{1,s} \hat c^\dag_{1,s}\hat c_{2,s} \hat c^\dag_{2,s}\hat c_{3,s} \hat c^\dag_{3,s}),
\end{eqnarray}
where $V_{123}^{(6)}=(V_{123}^{(6)})^*$. It is straightforward to see that above interaction is identical to a summation of two-body interactions and can not split the degeneracy of the ground states. Other possible 3-chain interactions include
\begin{eqnarray}
H'_{123}&=& \sum_s V_{123}^{(4)}\left[\left(c_{1,s}^\dag c_{2,s}^\dag c_{2,s}c_{3,s} + c_{1,s} c_{2,s} c_{2,s}^\dag c_{3,s}^\dag \right) + \mathrm{h.c.}\right]\\
&=&V_{123}^{(4)}(2c_{2,s}^\dag c_{2,s}-1) (c_{1,s}^\dag c_{3,s}+c_{3,s}^\dag c_{1,s}),
\end{eqnarray}
It can be easily checked that in the case (\ref{GS3_1}) the perturbation $H'_{123}$ have zero matrix elements in the ground-state subspace and hence can not split the degeneracy. The condition is similar in another case (\ref{GS3_2}). The difference is that, in case (\ref{GS3_2}), $H'_{123}$ mixes the ground states with higher energy states. For instance, the state $|01\rangle_1|01\rangle_2|10\rangle_3$ is mixed with $|11\rangle_1|01\rangle_2|00\rangle_3$ to lower its energy. At the same time the state $|10\rangle_1|10\rangle_2|01\rangle_3$ is mixed with $|00\rangle_1|10\rangle_2|11\rangle_3$ to lower the energy with the same amount. As a result, the two new states, as the new ground states, are still degenerate. Therefore, the topological properties of 3 chains are stable against interactions respecting the symmetry, and we need to investigate the 4-chain system.

Finally, for $N=4$, it turns out that we can find a path to smoothly reduce the ground state degeneracy to 1. It is easy to verify that degeneracy of the ground states can be reduced to 4 under two-body interaction $H_{\rm int}=H_{12}+H_{23}+H_{34}$ with $V_{12}=-V_{23}=V_{34}=-V_0<0$. These four ground states are given by
\begin{eqnarray}\label{eqn:GS4}
&&|01\rangle_{1}|01\rangle_{2}|10\rangle_{3}|10\rangle_{4}, \ \ |10\rangle_{1}|10\rangle_{2}|01\rangle_{3}|01\rangle_{4},\nonumber \\
&&|11\rangle_{1}|11\rangle_{2}|00\rangle_{3}|00\rangle_{4}, \ \ |00\rangle_{1}|00\rangle_{2}|11\rangle_{3}|11\rangle_{4},
\end{eqnarray}
with the energy equal to $-4V_0$. However, the above ground states can be further gapped out by taking into account the following interactions:
\begin{eqnarray}\label{4chain}
H_{1234}=\sum_s(V_{1234}^{(4)}c_{1,s}^\dag c_{2,s}^\dag c_{3,s}\ c_{4,s} + V_{1234}^{(4)}c_{1,s}\ c_{2,s}c_{3,s}^\dag c_{4,s}^\dag).
\end{eqnarray}
In the 4-dimensional Hilbert space spanned by the ground states in Eq.~\eqref{eqn:GS4}, the above interaction can be written in the matrix form
\begin{eqnarray}
{\cal H}_{1234}=V_{1234}^{(4)}\left(\begin{matrix}
0&0&1&1\\
0&0&1&1\\
1&1&0&0\\
1&1&0&0
\end{matrix}\right),
\end{eqnarray}
whose eigenvalues are $-2V_{1234}^{(4)}, 0, 0$, and $2V_{1234}^{(4)}$. We therefore obtain the single non-degenerate ground state with the energy $-2|V_{1234}^{(4)}|-4V_0$. This implies that under interaction the 4-chain system can be smoothly connected to a trivial phase without closing the bulk gap, and we therefore complete the proof that the $Z$ classification can be broken down to $Z_4$ under interactions.

\section{Particle fractionalization}

We prove in this section that each edge state leads to $1/2$-fractionalization. A convenient way is to consider the semi-infinite geometry which has the open boundary at $x=0$. We then calculate the particle number of the zero mode localized on this boundary. Note the total number of quantum states in the system is given by
\begin{eqnarray}\label{eqn:apparticle1}
N=\sum_{E<0}\langle\psi_E|\hat n_E|\psi_E\rangle+\sum_{E>0}\langle\psi_E|\hat n_E|\psi_E\rangle+\langle\psi_0|\hat n_E|\psi_0\rangle,
\end{eqnarray}
where we denote by $\hat n_E=\mathbb{I}$ the state number operator and $|\psi_E\rangle$ is the eigenstate with energy $E$. Since the Hamiltonian satisfies $\{{\cal H},\sigma_x\}=0$, the energy spectrum is symmetric. We have then $\sum_{E<0}\langle\psi_E|\hat n_E|\psi_E\rangle=\sum_{E>0}\langle\psi_E|\hat n_E|\psi_E\rangle$. It follows that
\begin{eqnarray}\label{eqn:apparticle2}
\sum_{E<0}\langle\psi_E|\hat n_E|\psi_E\rangle=\frac{1}{2}[N-\langle\psi_0|\hat n_E|\psi_0\rangle].
\end{eqnarray}
The particle number of the zero mode depends on its occupation. If the zero mode is unoccupied, the particle number of it is given by
\begin{eqnarray}\label{eqn:apparticle3}
n_0=\sum_{E<0}\bigr[\langle\psi_E|\hat n_E|\psi_E\rangle_1-\langle\psi_E|\hat n_E|\psi_E\rangle_0\bigr].
\end{eqnarray}
Here $\langle\rangle_1$ and $\langle\rangle_0$ represents the cases with one (topological phase) and zero (trivial phase) bound modes, respectively. Using the Eq.~\eqref{eqn:apparticle2} one finds directly
\begin{eqnarray}\label{eqn:apparticle3}
n_0=-\frac{1}{2}\langle\psi_0|\hat n_E|\psi_0\rangle=-\frac{1}{2}.
\end{eqnarray}
Similarly, if the zero mode is occupied, the particle number is $n_0=1/2$. It is trivial to know that this result can be applied to the case with two boundaries located far away from each other, say respectively at $x=0$ and $x=L$. Since the two zero modes are obtained independently, each of them carries $+1/2$ ($-1/2$) particle if it is occupied (unoccupied).

\section{Derivation of the RG equations}

In this section we provide the derivation of the one-loop RG equations. We consider that in Hamiltonian~\eqref{eqn:aptightbinding} only one Zeeman term, e.g. $\Gamma_y$ is nonzero (the case with $\Gamma_z\neq0$ can be studied in the similar way). We find it convenient to redefine that $\hat c_{j\downarrow}\rightarrow e^{i\pi x_j/a}\hat c_{j\downarrow}$ and rewrite the Hamiltonian~\eqref{eqn:aptightbinding} in the form: $H=-t_s\sum_{<i,j>,\sigma}\hat c_{i\sigma}^{\dag}\hat
c_{j\sigma}+H_\text{SO}+H_\text{Z}$, with $H_\text{SO}=t_{\rm so}^{(0)}\sum_j (-1)^j(c_{j\uparrow}^\dag c_{j+1,\downarrow}-c_{j\uparrow}^\dag c_{j-1,\downarrow}+\text{h.c.})$ and $H_\text{Z}=\Gamma_y\sum_j (-1)^j(ic_{j\downarrow}^\dag c_{j\uparrow}-\text{h.c.})$. The low-energy physics can be well captured by the continuum approximation ($x=ja$):
\begin{equation}
  c_{j\sigma}\approx \sqrt{a}[\psi_{R\sigma}(x)e^{ik_Fx}+\psi_{L\sigma}(x)e^{-ik_Fx}],
  \label{}
\end{equation}
with $k_F=\pi/2$. The continuum representation of the two mass terms are then given by:
\begin{equation}
  \begin{gathered}
	  H_\text{SO}\approx iu\int dx\,(\psi_{L}^\dag\sigma^x\psi_R-\psi_{R}^\dag\sigma^x\psi_{L}), \ \
	  H_\text{Z}=w\int d x\,(\psi_R^\dag \sigma^y\psi_L+\psi_L^\dag\sigma^y\psi_R).
  \end{gathered}
    \label{eqn:mass}
\end{equation}
Here $u=2t_{\text{so}}, w=\Gamma_y$. Using the standard bosonization formula $\psi_{rs}=\frac{1}{\sqrt{2\pi a}}e^{-\frac{i}{\sqrt{2}}[r\phi_\rho-\theta_\rho+s(r\phi_\sigma-\theta_\sigma)]}$ with $r=R,L$ and $s=\uparrow,\downarrow$,
we reach the bosonized Hamiltonian densities
\begin{equation}
	\begin{gathered}
	\mathcal{H}_\text{SO}=\frac{u}{\pi a}\sin\sqrt{2}\phi_\rho\cos\sqrt{2}\theta_\sigma, \ \
	\mathcal{H}_\text{Z}=\frac{w}{\pi a}\cos\sqrt{2}\phi_\rho\sin\sqrt{2}\theta_\sigma.
\end{gathered}
	\label{eqn:mass_bosonization}
\end{equation}

The allowed four-fermion interactions in the (unperturbed) Hubbard model at half-filling is highly constrained by the $\mathbb{SU}(2)\times\mathbb{SU}(2)$ symmetry. It is convenient to define current operators:
\begin{equation}
  J_{r}=\psi_{r\alpha}^\dag\psi_{r\alpha}, \vec{J}_r=\frac{1}{2}\psi_{r\alpha}^\dag\pmb{\sigma}_{\alpha\beta}\psi_{r\beta} , I_r=\psi_{r\alpha}\epsilon_{\alpha\beta}\psi_{r\beta}.
  \label{eqn:currents}
\end{equation}
Here $\epsilon$ is the fully anti-symmetric tensor.
The general form of four-fermion interaction is given by
\begin{equation}
  \mathcal{H}_\text{int}=2g_\sigma\vec{J}_R\cdot\vec{J}_L+\frac{g_{\rho}}{4}(I_R^\dag I_L+\text{h.c.}).
  \label{}
\end{equation}
We then derive the RG flow equations to understand the fate of the topological phase transition driven by the competition of the staggered Zeeman term and the spin-orbit coupling term. The tree level term can be easily read off from the scaling dimensions of $\mathcal{H}_\text{SO}$ and $\mathcal{H}_\text{Z}$ in their bosonization form~\eqref{eqn:mass_bosonization}, both of which are $(1+K_\rho)/2$. So it is necessary to go to the next order in the perturbative expansion. The one-loop order RG flow can be most easily derived by calculating the operator algebra of the various operators~\cite{Cardy, RG}.

To derive the RG equation, we consider the partition function in the Euclidean functional integral representation
\begin{equation}
  \mathcal{Z}=\int\mathcal{D}\overline{\psi}\mathcal{D}\psi e^{-S}.
  \label{}
\end{equation}
The Euclidean action $S=\int d x d \tau\, (\mathcal{H}_0+\mathcal{H}_1)$ where $\mathcal{H}_0=\sum_{\nu=\rho,\sigma}\frac{v_\nu}{2}[{K}_\nu(\partial_x\theta_\nu)^2\!+\!K_\nu^{-1}(\partial_x\phi_\nu)^2]$ is the unperturbed Gaussian part and the $\mathcal{H}_1=\mathcal{H}_\text{SO}+\mathcal{H}_\text{Z}$ is the perturbation.
To perform the RG, we expand the exponential to the second order in $\mathcal{H}_1$. Let us write $\mathcal{H}_1=\sum g_i\mathcal{O}_i$. Then the second order term is given by
\begin{equation}
  \frac{1}{2}g_ig_j\sum_{i,j} \int_{z,w}\langle \mathcal{O}_i(z)\mathcal{O}_j(w)\rangle\simeq\frac{1}{2}g_ig_j\sum_{i,j} \int_{z,w}\frac{c_{ijk}}{4\pi^2|z-w|^2}\mathcal{O}_k.
  \label{eqn:rg1}
\end{equation}
We have introduced complex coordinates $z, w$ where $z=v\tau-ix$. Here we assume the following operator product expansion (OPE)
\begin{equation}
  \mathcal{O}_i(z)\mathcal{O}_j(0)\sim \frac{c_{ijk}}{4\pi^2|z|^{2}}\mathcal{O}_k+\text{regular terms},
  \label{}
\end{equation}
which is sufficient for our purpose. The OPEs are valid when two points $z$ and $0$ are brought close together, as replacement within correlation functions.

At this point the cutoff prescription needs to be carefully specified. We will choose a short-distance cutoff $a$ in space, but none in imaginary time. For a rescaling factor $b$, we must then perform the integral
\begin{equation}
  I=\int_{a<|x|<ba}d x\int_{-\infty}^\infty d \tau \frac{1}{v^2\tau^2+x^2}=\frac{2\pi}{v}\ln b.
  \label{}
\end{equation}
Eq. \eqref{eqn:rg1} then becomes
\begin{equation}
  \frac{c_{ijk}}{4\pi v}g_ig_j\ln b \int \mathcal{O}_k.
  \label{}
\end{equation}
Upon re-exponentiating we obtain the one-loop RG equation
\begin{equation}
  \frac{dg_k}{dl}=-\frac{1}{4\pi v}\sum_{ij}c_{ijk}g_ig_j.
  \label{}
\end{equation}
Here $l=\ln b$.

Now let us be more specifit. Besides the current operators defined in \eqref{eqn:currents}, we also need to define
\begin{equation}
  M_i=\psi_R^\dag\sigma^i\psi_L,\: i=x,y.
  \label{}
\end{equation}
The fermionic field operators satisfy the following OPEs:
\begin{equation}
  \begin{gathered}
  \psi_{R\alpha}(z)\psi_{R\beta}^\dag(0)\sim\frac{\delta_{\alpha\beta}}{2\pi z}, \ \
   \psi_{L\alpha}(z)\psi_{L\beta}^\dag(0)\sim\frac{\delta_{\alpha\beta}}{2\pi z^*}.
  \end{gathered}
  \label{eqn:ope1}
\end{equation}
The OPEs of the currents and the mass $M_i$ can be calculated from \eqref{eqn:ope1} using Wick's theorem. Those between the currents are standard and can be found in [\onlinecite{RG}] which we do not duplicate. Below are the needed ones:
\begin{equation}
  \begin{gathered}
	M_i(z)\vec{J}_R\cdot\vec{J}_L(0)\sim\frac{1}{16\pi^2|z|^2}M_i(z),\ \ M_i^\dag(z)\vec{J}_R\cdot\vec{J}_L(0)\sim\frac{1}{16\pi^2|z|^2}M_i^\dag(z)\\
	M_i(z)[I_RI_L^\dag+\text{h.c.}](0)\sim -\frac{2}{4\pi^2|z|^2}M_i^\dag(z),\ \ M_i^\dag(z)[I_RI_L^\dag+\text{h.c.}](0)\sim -\frac{2}{4\pi^2|z|^2}M_i(z).
  \end{gathered}
  \label{}
\end{equation}
Applying the above formalism to the model at hand, we find
\begin{equation}
  \begin{split}
	  \frac{du}{dl}&=\frac{3-K_\rho}{2}u-\frac{g_{\rho}u}{4\pi v_F}+\frac{g_\sigma u}{4\pi v_F},\\
  \frac{dw}{dl}&=\frac{3-K_\rho}{2}w+\frac{g_{\rho}u}{4\pi v_F}+\frac{g_\sigma u}{4\pi v_F},\\
  \frac{dg_{\sigma}}{dl}&=\frac{1}{\pi v_F}g_\sigma^2, \ \
  \frac{dg_{\rho}}{dl}=\frac{1}{\pi v_F}g_{\rho}^2.
  \end{split}
  \label{}
\end{equation}
If we take the interaction in the lattice model to be an on-site Hubbard form: $H_U=U\sum_i n_{i\uparrow}n_{i\downarrow}$,
the bare values of the coupling coefficients are
\begin{equation}
  g_\rho=U, g_\sigma=-U.
  \label{}
\end{equation}
Then for the repulsive interaction $U>0$, the coupling parameter $g_\sigma$ of the spin sector is marginally flows to zero, while $g_\rho$ is marginally relevant.

\section{Topological edge spins on soft boundaries}

\begin{figure}[t]
\includegraphics[width=0.55\columnwidth]{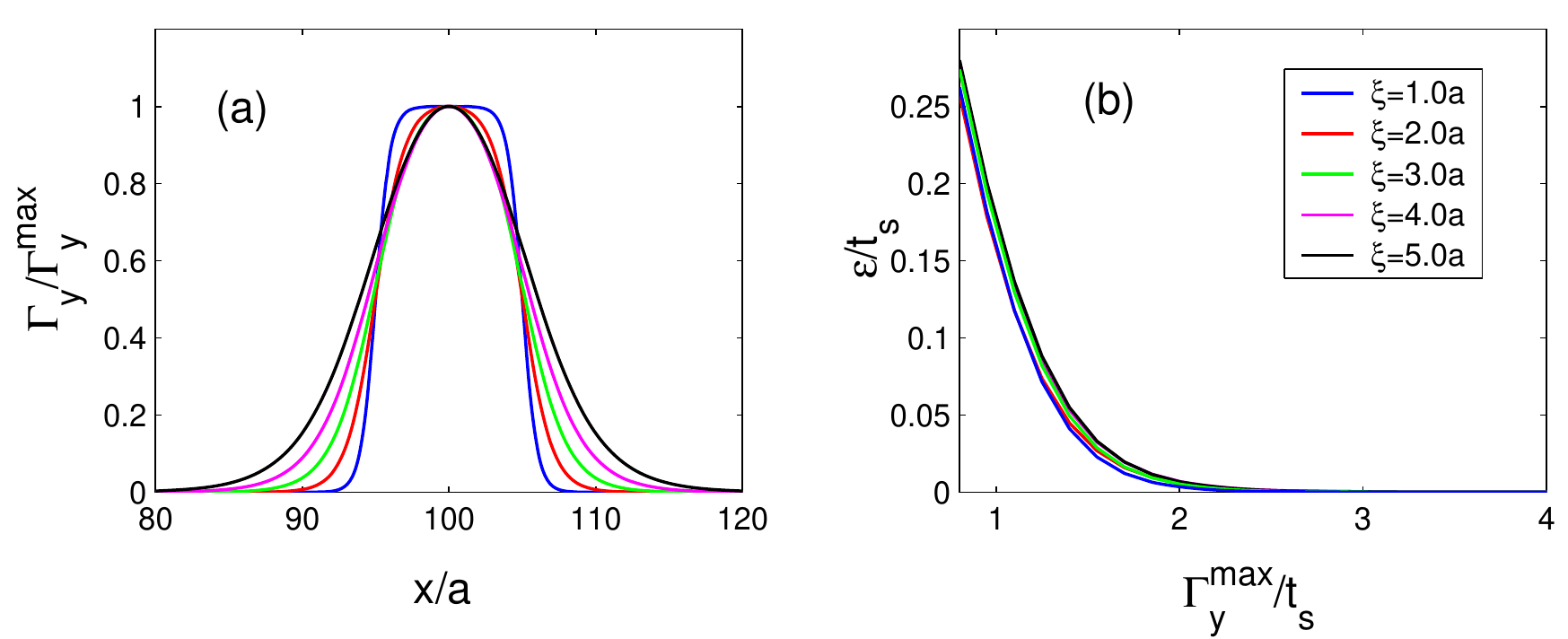} \caption{(Color online) (a) Mass domain with soft domain walls; (b) The couplings between the edge spins localized on the left and right domain walls, where $\Gamma_y^{\rm max}$ is the maximum value of $\Gamma_y(x)$ and it has $\Gamma_y^{\rm max}=\Gamma_y(L/2)$.}
\label{splitting_appendix}
\end{figure}
The appearance of the edge modes localized in the boundary between trivial insulating and topological insulating regions is protected by the nontrivial winding number in the bulk, and is independent of the details of the boundary. The different configurations of the mass domain walls may at most quantitatively affect the coupling between the localized edge spins, as numerically shown in Fig.~\ref{splitting_appendix}, where we consider the mass domain with soft boundaries, described by [Fig.~\ref{splitting_appendix}(a)]
\begin{equation}
\Gamma_y(x)=\Gamma_0\biggr\{-\tanh\bigr[\frac{x-L/2+5a}{\xi}\bigr]+\tanh[\frac{x-L/2-5a}{\xi}\bigr]\biggr\}.
  \label{}
\end{equation}
Here $L$ is the length of the lattice and $\xi$ represents the domain wall length. The distance between the left and right domain walls is $10a$. In Fig.~\ref{splitting_appendix}(b) we have shown that by varying the domain length from $\xi=a$ (the profile is close to step functions) to $\xi=5a$ (the profile is close to a Gaussian function), the coupling slightly increases for fixed $\Gamma_y^{\rm max}$ [$=\Gamma_y(L/2)$]. The increasing coupling is because the wave-function overlapping of the localized spins is enhanced for wider domain walls, which can lead to a slightly larger spin Rabi-oscillation frequency, and a shorter manipulation time for the TSQs.

\end{appendix}

\end{widetext}


\begin{thebibliography}{99}

\bibitem{QHE1} K.V. Klitzing, G. Dorda, and M. Pepper, Phys. Rev. Lett. {\bf 45}, 494 (1980); D. C. Tsui, H. L. Stormer, and A. C. Gossard, Phys. Rev. Lett. {\bf 48}, 1559 (1982).

\bibitem{TP} F. D. M. Haldane, Phys. Rev. Lett. {\bf 50}, 1153 (1983); Phys. Lett. {\bf 93}, 464 (1983);  N. Read and D. Green, Phys. Rev. B {\bf 61}, 10267 (2000);  A. Y. Kitaev, Physics-Uspekhi {\bf 44}, 131 (2001).

\bibitem{QSH1} C. L. Kane and E. J. Mele, Phys. Rev. Lett. {\bf 95}, 146802 (2005); B. A. Bernevig,
T. L. Hughes and S. -C. Zhang, Science, {\bf 314} 1757 (2006); L. Fu, C. L. Kane, and E. J. Mele, Phys. Rev. Lett. {\bf 98}, 106803 (2007);
J. E. Moore and L. Balents, Phys. Rev. B {\bf 75}, 121306 (R) (2007); R. Roy, {\it ibid} {\bf 79}, 195322 (2009).

\bibitem{QSH2} M. Z. Hasan and C. L. Kane, Rev. Mod. Phys. {\bf 82}, 3045 (2010);
X. -L. Qi and S. -C. Zhang, {\it ibid} {\bf 83}, 1057 (2011).

\bibitem{A-Z} A. Altland and Martin R. Zirnbauer, Phys. Rev. B {\bf 55}, 1142 (1997).

\bibitem{Wen} Z. C. Gu and X. G. Wen,  Phys. Rev. B {\bf 80}, 155131 (2009); X. Chen, Z. -C. Gu, and X. -G. Wen, {\it ibid} {\bf 82}, 155138 (2010); X. -G. Wen, {\it ibid} {\bf 85}, 085103 (2012).

\bibitem{Ryu} Shinsei Ryu, Andreas Schnyder, Akira Furusaki, Andreas Ludwig, New J. Phys. {\bf 12}, 065010 (2010).

\bibitem{SPT}  L. Fidkowski and A. Kitaev, Phys. Rev. B {\bf 81}, 134509
(2010); A. M. Turner, F. Pollmann, and E. Berg, {\it ibid}
{\bf 83}, 075102 (2011); L. Fidkowski and A. Kitaev, {\it ibid} {\bf 83}, 075103
(2011).

\bibitem{TQC} Y. Kitaev, Ann. Phys. 303, 2 (2003); S. Das Sarma, M. Freedman, and C. Nayak, Phys. Rev.
Lett. 94, 166802 (2005); C. Nayak et al., Rev. Mod. Phys. {\bf 80}, 1083 (2008).

\bibitem{Liu} X. -J. Liu, M. F. Borunda, X. Liu, and J. Sinova, Phys. Rev. Lett. {\bf 102}, 046402 (2009).

\bibitem{Lin} Y.-J. Lin, K. Jim\'{e}nez-Garc\'{i}a, and  I. B. Spielman, Nature {\bf 471}, 83 (2011).

\bibitem{Chapman} M. Chapman and C. S\'{a} de Melo, Nature {\bf 471}, 41 (2011).

\bibitem{Wang} P. Wang et al., Phys. Rev. Lett. {\bf 109}, 095301 (2012).

\bibitem{MIT} L. W. Cheuk, A. T. Sommer, Z. Hadzibabic, T. Yefsah, W. S. Bakr, and M. W. Zwierlein, Phys. Rev. Lett. {\bf 109}, 095302 (2012).

\bibitem{Pan} Jin-Yi Zhang et al., 
Phys. Rev. Lett. {\bf 109}, 115301 (2002).

\bibitem{SOC} J. Larson, J.-P. Martikainen, A. Collin, and E. Sjoqvist, Phys. Rev. A {\bf 82}, 043620 (2010); D. Sokolovski and E. Ya. Sherman, {\it ibid} {\bf 84}, 030101(R) (2011); J. D. Sau, R. Sensarma, S. Powell, I. B. Spielman, and S. Das Sarma, Phys. Rev. B {\bf 83}, 140510(R) (2011); T. Ozawa and G. Baym, Phys. Rev. Lett. {\bf 109}, 025301 (2012); G. I. Martone, Y. Li, L. P. Pitaevskii, and S. Stringari, arXiv:1207.6804 (2012).

\bibitem{Liu1} X. -J. Liu, X. Liu, L. C. Kwek and C. H. Oh, Phys. Rev. Lett. {\bf 98}, 026602 (2007); Phys. Rev. B \textbf{79}, 165301 (2009).

\bibitem{Wu} C. Wu, Phys. Rev. Lett. {\bf 101}, 186807 (2008); X.-J. Liu, X. Liu, C. Wu, and J. Sinova, Phys. Rev. A {\bf 81}, 033622 (2010); Y. Yu and K. Yang, Phys. Rev. Lett. {\bf 105}, 150605 (2010).

\bibitem{Goldman} N. Goldman, I. Satija, P. Nikolic, A. Bermudez, M.A. Martin-Delgado, M. Lewenstein, and I. B. Spielman, Phys. Rev. Lett. {\bf 105}, 255302 (2010); N. Goldman, J. Beugnon, and F. Gerbier, {\it ibid} {\bf 108}, 255303 (2012).

\bibitem{Li} X. Li, E. Zhao, and V. W. Liu, arXiv:1205.0254 (2012). 

\bibitem{Wuming} G. Liu, S. -L. Zhu, S. Jiang, F. Sun, and W. M. Liu, Phys. Rev. A {\bf 82}, 053605 (2010); F. Mei, S. -L. Zhu, Z. -M. Zhang, C. H. Oh, and N. Goldman, {\it ibid} {\bf 85}, 013638 (2012).

\bibitem {Chuanwei1} C. Zhang, S. Tewari, R. Lutchyn, and S. Das Sarma, Phys. Rev. Lett. {\bf 101}, 160401 (2008);
Y. Zhang, L. Mao, C. Zhang, {\it ibid} {\bf 108}, 035302 (2012).

\bibitem{Sato} M.Sato, Y. Takahashi, and S. Fujimoto, Phys. Rev. Lett. 103 020401 (2009).

\bibitem{zhu1} S. -L. Zhu, L. B. Shao, Z. D. Wang, and L. -M. Duan, Phys. Rev. Lett. {\bf 106}, 100404 (2011); W. Yi and G.-C. Guo, Phys. Rev. A {\bf 84}, 031608(R) (2011); L. He and X. -G. Huang, Phys. Rev. Lett. {\bf 108}, 145302 (2012); Phys. Rev. B {\bf 86}, 014511 (2012).

\bibitem{Seo} Kangjun Seo, Li Han, and C. A. R. S\'{a} de Melo, Phys. Rev. Lett. {\bf 109}, 105303 (2012).

\bibitem{Kraus} C. V. Kraus, S. Diehl, P. Zoller, and M. A. Baranov, arXiv:1201.3253; S. Nascimb\`{e}e, arXiv:1210.0687.

\bibitem{RMPcoldatom} I. Bloch, J. Dalibard, and W. Zwerger, Rev. Mod. Phys. \textbf{80}, 885 (2008).

\bibitem{SI} See Supplementary Material for more details.

\bibitem{hardwall} T. P. Meyrath, F. Schreck, J. L. Hanssen, C.-S. Chuu, and M. G. Raizen,
Phys. Rev. A {\bf 71}, 041604(R) (2005).

\bibitem{note0} Note the degeneracy and spin-polarization of the edge modes obtained by F. Mei et al. in Ref.~\cite{Wuming} are not symmetry protected and not stable against local perturbations. Therefore such edge modes cannot form TSQ.

\bibitem{book} T. Giamarchi, Quantum physics in one dimension (Oxford University Press, 2004).



\bibitem{RG} L. Balents and M. P. A. Fisher, Phys. Rev. B {\bf 53}, 12133 (1996).

\bibitem{Schlosser} N. Schlosser, G. Reymond, I. Protsenko, and P. Grangier, Nature, {\bf 411}, 1024 (2001).

\bibitem{note} We neglect the couplings between $|\psi_\pm\rangle$ and bulk states $|u_k\rangle$, which is valid when $|\langle\psi_\pm|\partial_t|u_k\rangle|\ll E_g$ during the manipulation.

\bibitem{Volz} J. Volz, R. Gehr, G. Dubois, J. Est\`{e}ve, and J. Reichel, Nature {\bf 475}, 210 (2011).

\bibitem{SSR} W. S. Bakr, J.I. Gillen, A. Peng, S. Foelling, and M. Greiner, Nature {\bf 462}, 74 (2009); M. Karski, L. F\"{o}rster, J. M. Choi, W. Alt, A. Widera, and D. Meschede, Phys. Rev. Lett. {\bf 102}, 053001 (2009).


\end{thebibliography}

\begin{thebibliography}{99}

\bibitem{Cardy} J. Cardy, Scaling and Renormalization in Statistical Physics (Cambridge University Press, 1996)
\bibitem{RG} L. Balents and M. P. A. Fisher, Phys. Rev. B {\bf 53}, 12133 (1996).

\end{thebibliography}
\end{document}